\documentclass[twocolumn,nofootinbib,aps,floatfix,superscriptaddress,preprintnumbers]{revtex4}

\usepackage{graphicx}% Include figure files

\begin{document}

%%%%%%%%%%%%%%%%%%%%%%%%%%%%%%%%%%%%%%%%%%%%%%%%%%%%%%%%%%%%%%%%%%%%%%%%%
%                      BEGINNING OF DOCUMENT                            %
%%%%%%%%%%%%%%%%%%%%%%%%%%%%%%%%%%%%%%%%%%%%%%%%%%%%%%%%%%%%%%%%%%%%%%%%%

\preprint{
\vbox{
%\hbox{CERN-OPEN-2007-xxx}
\hbox{CPT-P08-2007}
%\hbox{LAL 07-xxx}
\hbox{LAPP-EXP-2007-01}
}}

\vspace*{1mm}

\title{\boldmath Reply to: ''Improved Determination of the CKM Angle $\alpha$ from $B \to \pi \pi$ decays''}

\author{J.~Charles}
\affiliation{CPT, Luminy Case 907, F-13288 Marseille Cedex 9, France}

\author{A.~H\"ocker}
\affiliation{CERN, CH-1211 Geneva 23, Switzerland}

\author{H.~Lacker}
\affiliation{TU Dresden, IKTP, D-01062 Dresden, Germany}

\author{F.R. Le Diberder}
\affiliation{LAL, Universit\'e Paris-Sud 11, CNRS/IN2P3, B\^at. 200, BP 34, F-91898 Orsay Cedex, France}

\author{S. T'Jampens}
\affiliation{LAPP, Universit\'e de Savoie, CNRS/IN2P3, 9 Chemin de Bellevue, BP 110, F-74941 Annecy-le-Vieux Cedex, France}

\date{\today}

\begin{abstract}
In reply to Ref.~\cite{nimportenawak} we demonstrate why the arguments
made therein do not address the criticism exposed in 
Ref.~\cite{bayestrouble} on the fundamental shortcomings of the Bayesian
approach when it comes to the extraction of parameters of Nature from
experimental data. As for the isospin analysis and the CKM angle $\alpha$
it is shown that the use of uniform priors for the observed quantities in
the Explicit Solution parametrization is equivalent to a frequentist
construction resulting from a change of variables, and thus relies neither
on prior PDFs nor on Bayes' theorem. This procedure provides in this
particular case results that are similar to the Confidence Level
approach, but the treatment of mirror solutions remains incorrect and it
is far from being general. In a second part it is shown that important
differences subsist between the Bayesian and frequentist approaches, when
following the proposal of Ref.~\cite{nimportenawak} and inserting
additional information on the hadronic amplitudes beyond isospin
invariance. In particular the frequentist result preserves the exact
degeneracy that is expected from the remaining symmetries of the problem
while the Bayesian procedure  does not. Moreover, in the
Bayesian approach reducing inference to the  68\% or 95\% credible
interval is a misconception of the meaning of the posterior PDF, which in
turn implies that the significant dependence of the latter to the chosen
parametrization  cannot be viewed as a minor effect, contrary to the claim
in  Ref.~\cite{nimportenawak}. \end{abstract}

\maketitle

%
% -------------------------------------------------------------------------------------
%

\section{Introduction}

In  Ref.~\cite{bayestrouble} we have shown through the example of the
extraction of the Cabibbo-Kobayashi-Maskawa (CKM) angle $\alpha$ from
$B\to\pi\pi,\,\rho\rho$ decays
that the Bayesian treatment as done, e.g., by the UTfit 
collaboration~\cite{UTfitpapers}, suffers from major difficulties. The problems
we have found are related to the presence of unknown free parameters
which should actually be constrained by the data. We have shown that
the results of the Bayesian analysis depend on the priors and the chosen
parameterization in an uncontrollable manner, and may even diverge in some
cases. Furthermore we have demonstrated that the priors cannot always be
specified in a consistent way with respect to the symmetries of
the problem, which results in the present example in  an incorrect
description of the $\alpha\to 0$ limit.

The authors of Ref.~\cite{nimportenawak} replied to our criticism.
In the first part of their paper they agree with us on the dependence of
the analysis with respect to the choice of priors and parameterization,
even when the data are sufficiently precise so that the posterior PDFs
are contained in the corresponding prior ranges.
Ref.~\cite{nimportenawak} claims that this should not be viewed as a
fundamental drawback of the Bayesian treatment, because in the absence of
additional information on the parameters one should use uniform priors on the
quantities that are directly measured (what we have called the "Explicit
Solution" parameterization in~\cite{bayestrouble}). We emphasize in the
following that this recipe can only be (approximately) justified within
the framework of classical statistics and is thus not actually Bayesian. 
Moreover, it is far from being general.

In the second part of Ref.~\cite{nimportenawak} it is argued that in addition to
the isospin analysis one may use bounds on the magnitude of the fit parameters
related to hadronic matrix elements, and when taken into account the Bayesian 
treatment behaves more reliably. While this new recommendation may appear 
physically sound, the problem reported in~\cite{bayestrouble} uses the 
original isospin analysis~\cite{GL}, where the hadronic parameters are unknown
and only isospin symmetry is used. Changing the hypotheses cannot be 
considered as a satisfying answer to our findings: indeed the validity 
of the statistical approach must be proven independently of the problem at 
hand. Despite this inconsistency, we accept the new proposal and show below that 
even with the use of this additional information one finds important
quantitative and qualitative 
differences between the frequentist and the Bayesian results.

In any case, the new recommendations of the UTfit collaboration for
treating problems with {\em a priori} free unknown parameters differ
quite importantly from their previous publications, which implies
that the results published therein are obsolete. We are therefore expecting 
new analysis results for the angles $\alpha$ from $B\to\rho\pi$ and
$\gamma$ from charmful modes, 
for which similar difficulties can be apprehended.

We conclude that Ref.~\cite{nimportenawak} does not abrogate our criticism, 
because the only known examples where the Bayesian
treatment leads to a numerically reasonable result 
do not provide any general argument in favor of the underlying
hypotheses. In contrast, the frequentist approach can handle many
types of situations smoothly, independently of whether
there are free parameters and/or theoretical uncertainties.

\section{The original isospin analysis}

The term "isospin analysis" (which in Ref.~\cite{nimportenawak} is replaced by 
``minimal assumptions'') refers to the original paper by Gronau and London~\cite{GL}. 
Within this framework,  only isospin symmetry is used to parametrize the 
decay amplitudes. Although it is
not an exact symmetry, its accuracy in $B\to\pi\pi$ is expected to be of
the order of 1\%~\cite{GronauZupan} (once the known dominant contribution
from electroweak penguins has been taken into account) and is thus of
phenomenological relevance. Neither the Standard Model nor isospin symmetry
imply anything on the ``natural'' scale of the hadronic amplitudes.\footnote
{
   Note in passing that  isospin symmetry was historically found before
   the development of  QCD as the theory of the strong interactions, and the
   associated notion of a fundamental hadronic scale.
}

In Ref.~\cite{bayestrouble} we showed that the Pivk-Le Diberder (PLD) and
Explicit Solution (ES) parameterizations give the most reasonable answer
for the Bayesian treatment, because the corresponding parameters are
close or identical to the measured quantities. This is confirmed (for
the ES parameterization) by the
authors of Ref.~\cite{nimportenawak}, who thus agree  on the fact
that
other parameterizations lead to unstable results. The better behavior of
the ES description (which follows from our detailed analysis and should
be acknowledged as such) can be easily understood as follows. Let us stick 
to the Gaussian case and first assume that there is no degeneracy so that 
the angle $\alpha$ is given by a known single-valued function of
the observables. One can construct a sampling distribution for $\alpha$ as
defined by its analytical expression by randomizing  the
observables  according to  Gaussian PDFs where the
true (unknown) central values and standard deviations have been replaced
by their
corresponding measured estimators. It is then straightforward to show that
the position of the peak and its width are \textit{consistent estimators}
of, respectively, $\alpha$ and its uncertainty. This procedure does
not need to define priors nor integration on the true values of the
parameters, and does not rely on Bayes' theorem: it just follows from
standard rules of classical statistics.

Technically this procedure can be reproduced by applying the ``Bayesian'' 
treatment of Ref.~\cite{firstUTfit} to the ES parameterization~\cite{bayestrouble} 
with uniform priors.  Because for Gaussian measurements the standard deviation of a quantity 
is equal to the 68\% CL frequentist error, it is not a surprise to have
similar numbers in the ES parameterization between the ``Bayesian'' and
frequentist treatments. The remaining differences, namely the fact that
one finds exactly degenerate mirror solutions only in the frequentist
approach, mainly come from the integration (logical
AND) over mirror solutions in the randomization treatment, which results
from an unacceptable interpretation of physical constants that could
simultaneously take
different values~\cite{bayestrouble}.

Besides, the Explicit Solution trick is not general enough to treat with ``minimal
assumptions'' all the interesting physical problems. Even when there is
the same number of parameters as the number of independent observables,
there is no guarantee that the theoretical equations can be inverted to
express the parameters as functions of the observables. The
Gronau-London isospin analysis is quite special in this respect. More
importantly, most of the time one has overconstrained systems, that is one
has more observables than parameters. In this case no ``Explicit
Solution'' parameterization exists, and no general probability argument 
would help to find what is the most ``natural'' parameterization for 
a Bayesian treatment with uniform priors.

To close this Section we  point out two misconceptions in Ref.~\cite{nimportenawak}. 
The first one is related to the claim that the 68\% or 95\% confidence level 
or credible intervals are more meaningful than the full  curve. The
fact that experimental measurements are often summarized as one- or
two standard deviation intervals is just a matter of convention and if the
curve has a complicated shape, it cannot be reduced to a one or two-number
description without loss of information.\footnote
{
   ``{\em Posterior probability distributions provide the full description of our 
   state of knowledge about the value of the quantity. In fact, they allow 
   us to calculate all probability intervals of  interest. Such intervals 
   are also called credible intervals.[...] we emphasize that the full 
   answer is given by the posterior distribution, and reporting only these 
   summaries in the case of complex distributions (e.g. multimodal and/or 
   asymmetrical PDFs) can be misleading, because people tend to think of a 
   Gaussian model if no further information is provided.}'' (G.~d'Agostini, 2003~\cite{dago}).
} 
In particular for the present case, we emphasize the importance of having exactly 
degenerate peaks in the angle $\alpha$ (as are reproduced by the constrained 
frequentist fit, see below) to be in agreement with isospin symmetry no matter 
the values of the 68\% or 95\% error intervals.

The second misconception is related to the above misunderstanding of the meaning 
of the degeneracy that is intrinsic to the isospin analysis. The fact that
with the new data from Summer 2006~\cite{HFAG} one sees only four peaks
instead of eight as a function of $\alpha$ is not related to the
improved measurements, but to the fact that now the experimental central
values are slightly outside the physical region that is defined by the
isospin symmetry (which is related to the property that both the three
amplitudes and their $CP$ conjugates must form
a triangle in the complex plane). Most presumably the data will evolve in
such a way that the eightfold ambiguity reappears, which is the general
case
in the mathematical sense, and not a ``fortuitous accident'' as
claimed in~\cite{nimportenawak}.
Would a different ambiguity pattern be confirmed by better data, then
either the Standard Model or the isospin
symmetry should be questioned.

\section{Adding external information}
The authors of Ref.~\cite{nimportenawak} recommend that the ``default''
plot summarizing the present constraints on the angle $\alpha$ coming
from $B\to\pi\pi$ decays should take into account additional theoretical
and phenomenological information. The physical arguments that are
presented in favor of such an approach are perfectly legitimate and we
do not object them. However the problem
that was studied in detail in~\cite{bayestrouble} is not this one, but
rather the original isospin analysis. Let us
recall the advantages of performing the analysis assuming only isospin
symmetry.
\begin{itemize}
\item Neither isospin symmetry nor $B\to\pi\pi$ experimental measurements
 give any insight on the typical  scale of the hadronic amplitudes;
thus it is a natural choice for the experimental collaborations to present
their results in a way that is independent of possible assumptions on the
hadronic amplitudes.
\item The original paper~\cite{GL} only assumed isospin symmetry.
\item One can think of many relevant physical problems containing
completely free unknown parameters, the typical scale of which is not
controlled even as an order of magnitude. This is the case,
\textit{e.g.}, of general Beyond the Standard Model scenarios where one
parametrizes new arbitrary contributions with a few quantities that 
\textit{a priori} can take any value. The original isospin analysis is
thus a pedagogical example that exhibit many features that would appear
in more general situations.
\end{itemize}
Having this in mind we can now examine the new proposal of
Ref.~\cite{nimportenawak}. First we remark that the statement made
in~\cite{nimportenawak} according to which one can get new constraints on
$\alpha$, and eventually lift the degeneracies and suppress the solution at
$\alpha\rightarrow 0$, provided one uses additional theoretical and
phenomenological knowledge, is obviously trivial and was the guideline of, 
e.g., Section VI.1 of Ref.~\cite{ThePapII} and many similar studies in
the past.

Second we stress that additional information is
ill-defined. The assumptions made in~\cite{nimportenawak} are equivalent
to an
analysis with a finite theoretical error on each  parameter (except the
one that is scanned, here the angle $\alpha$), such as the ``historical''
determination of the Unitarity Triangle~\cite{firstUTfit}. It is well
known that in general there is no unambiguous definition of the meaning
of a theoretical error, and when such uncertainties are present the
Bayesian and frequentist
methods cannot be compared in a rigorous way. The authors of
Ref.~\cite{nimportenawak} take as a strong argument in favor of their
approach the fact that their results are weakly dependent of the assumed
order of magnitude for the amplitudes. We do not find this convincing, but
rather would like to take the following two limits,
 which to our knowledge are the only ones where
the concept of a theoretical error becomes perfectly well
defined.
\begin{itemize}
\item Theory errors go to zero, \textit{i.e.} the corresponding
parameters are fixed constants. In this case it is trivial to show that
both frequentist and Bayesian approaches lead to equivalent numerical
results.
\item Theory errors go to infinity, \textit{i.e.} the corresponding
parameters are completely free unknowns; this is the case that is
discussed in the previous section and where the Bayesian treatment in
a generic parameterization simply fails.
\end{itemize}

Hence, only in the very specific case where both experimental and
theoretical
uncertainties are ``sufficiently'' reduced, and there is no free
parameter, the numerical comparison of the Bayesian
treatment  with the frequentist
classical
approach may become meaningful,\footnote
{
   Trying to validate the Bayesian treatment by numerical comparison with
   the frequentist approach, without any internal consistency check, remains
   a quite peculiar working model.
} 
and in the Bayesian case one might expect a ``reasonable'' stability
with respect to  priors and parameterization. This stability, however, must be
extensively checked case-by-case, which has not been done in the previous
publications by the UTfit collaboration~\cite{UTfitpapers}.
 
\begin{figure}[tbp]
\begin{center}
\includegraphics[width=0.4\textwidth]{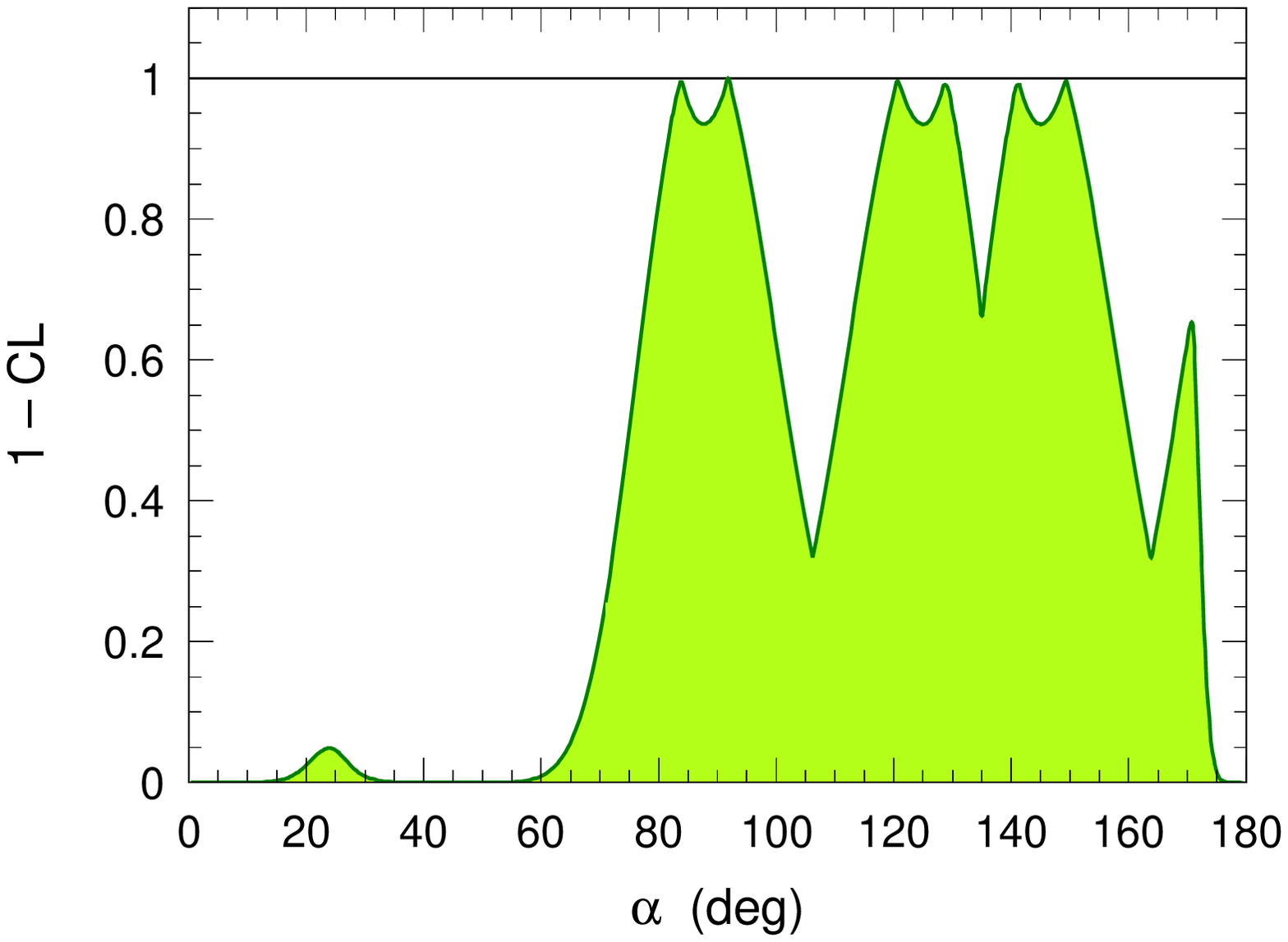}\hfill
\includegraphics[width=0.4\textwidth]{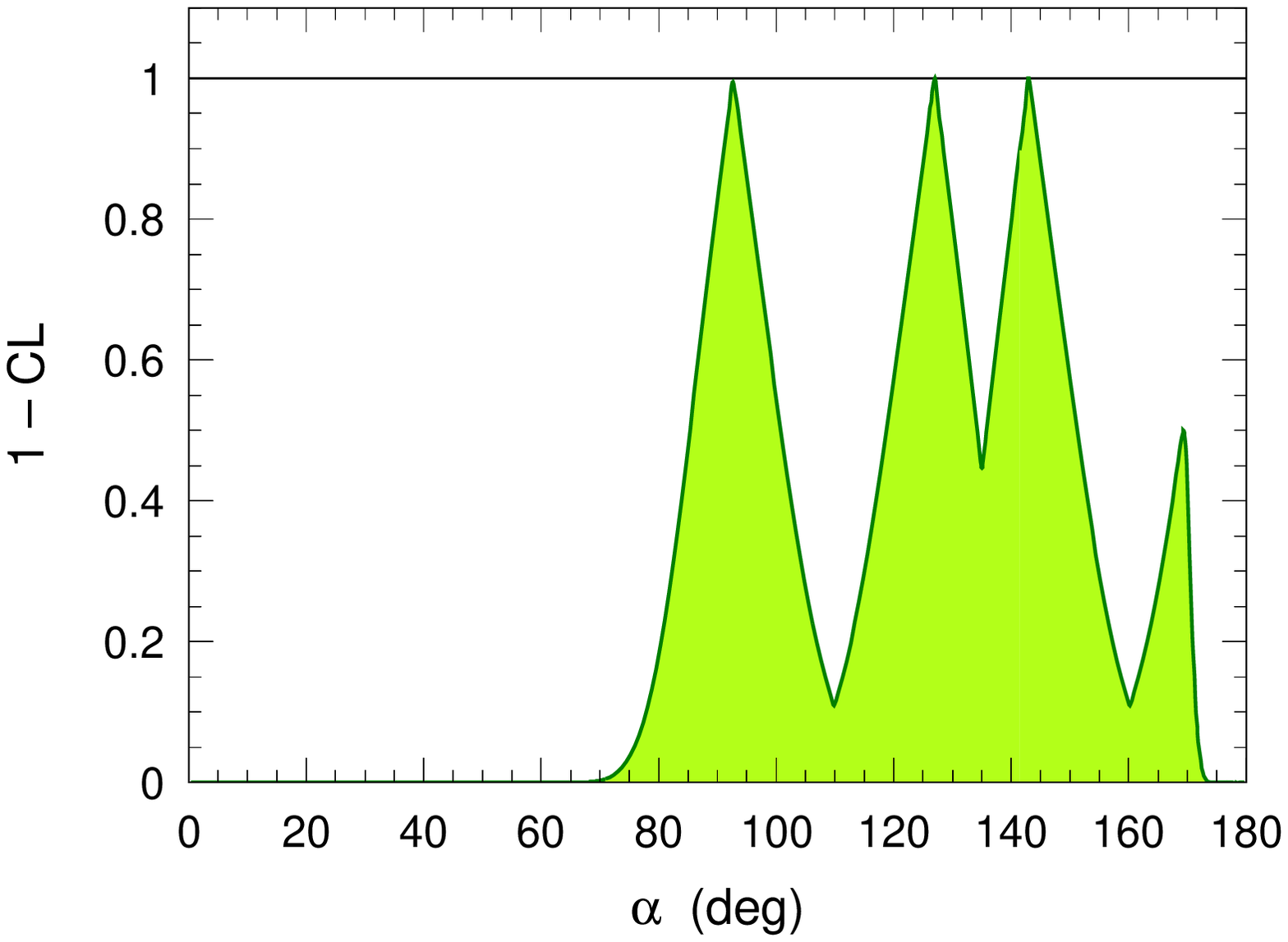}
\end{center}
\vspace{-0.3cm}
\caption{ Frequentist confidence level as a function of the  angle
$\alpha$, taking into account the following
constraints~\cite{nimportenawak}: $|T^{+-}|\le 10$, $|T^{00}|\le 10$ and
$|P|\le 2.5$ (natural
units), for $B\to\pi\pi$ decays. Upper plot: data that correspond to the plots
in Ref.~\cite{bayestrouble}, Lower plot: new data that correspond to Fig.~5
in~\cite{nimportenawak}.
\label{constrainedfit}}
\end{figure}

\begin{table}[tbp]
\caption{For each of the  peaks in the  plots of
Fig.~\ref{constrainedfit}
are shown the corresponding values for the parameters as found by direct
analytical calculation or by the numerical fit. The first six solutions
correspond to the upper plot, while the three others
correspond to the lower plot
\label{param}}
\centering
\vspace{0.2cm}
\setlength{\tabcolsep}{0.6pc}
{\normalsize
\begin{tabular*}{\columnwidth}{@{\extracolsep{\fill}}lcccc}\hline
%\begin{tabular}{lcccc} \hline
\hline
Solution & $\alpha$      & $|T^{+-}|$  & $|T^{00}|$  & $|P|$ \\
\hline
1        & $83.7^\circ$  & $1.74$      & $1.07$      & $0.763$ \\
2        & $91.8^\circ$  & $1.73$      & $1.28$      & $0.559$ \\
3        & $120.7^\circ$ & $2.01$      & $0.960$     & $0.650$ \\
4        & $128.8^\circ$ & $2.21$      & $0.600$     & $0.973$ \\
5        & $141.2^\circ$ & $0.960$     & $2.07$      & $1.21$  \\
6        & $149.3^\circ$ & $1.18$      & $2.16$      & $1.09$  \\
\hline
1        & $92.7^\circ$  & $1.71$      & $1.19$      & $0.634$ \\
2        & $126.9^\circ$ & $2.14$      & $0.722$     & $0.791$ \\
3        & $143.1^\circ$ & $1.17$      & $1.99$      & $1.05$  \\
\hline
\end{tabular*}}
\end{table}

Fig.~\ref{constrainedfit}
shows the results of a constrained frequentist fit that takes into
account the same information on the hadronic parameters as in Fig.~5
of~\cite{nimportenawak}, namely that the amplitudes verify the following
bounds: $|T^{+-}|\le 10$, $|T^{00}|\le 10$ and $|P|\le 2.5$ (in natural
units). In particular these bounds suppress the parameter configurations with
$\alpha\to 0$, as emphasized in~\cite{nimportenawak}.\footnote
{
   Without these bounds, arbitrary small, but finite, values for $\alpha$
   have a good confidence level because large, but finite, values for the
   hadronic amplitudes can generate the observed $CP$ violation. However 
   the $CP$-conserving value $\alpha=0$ is strongly disfavored in the Standard
   Model, so that the confidence level is discontinuous at this point, as
   explained in~\cite{bayestrouble}. Contrary to the claims in~\cite{nimportenawak},
   these results are  in perfect agreement with quantum field
   theory and do not violate any fundamental principle.
}
Indeed this curve is not too different from the one that corresponds to
the Bayesian
treatment, however it appears clearly that the frequentist analysis, in contrast 
to the Bayesian one, respects the exact degeneracy between the mirror
solutions that survive the phenomenological bounds, as requested
by the
symmetries of the problem. 
Table~\ref{param}
shows the parameter values  that correspond to each of the  peaks in the
 plots of Fig.~\ref{constrainedfit}, as obtained analytically from the
exact
formulas or numerically from the  fit.\footnote
{
   Using new data (lower plot in Fig.~\ref{constrainedfit}), the exact 
   degeneracy of the constrained fit is only threefold because the central 
   values are outside the physical region; again the  parameter values that
   correspond to the peaks exactly reproduce the same observable
   values, and all satisfy the phenomenological bounds on the amplitudes, so
   that there is no possibility to tell which solution is to be preferred.
}
Since these parameter configurations all lead to the very
same observables and all verify the phenomenological bounds advocated in
Ref.~\cite{nimportenawak}, there is no way to tell which configuration is
``more probable''. In turn, this means that the Bayesian treatment lifts
the degeneracy only because of the unphysical marginalization over the
hadronic
parameters, and that reducing the output information to the 68\% or 95\%
probability intervals is very misleading. From Table~\ref{param} one sees
that to lift the degeneracy
between the remaining solutions, even partially, one would need to know
the amplitudes with a relative uncertainty of better than 100\%, which is
feasible using more dynamical approaches to hadronic elements, but which would
request far more than just isospin symmetry and orders of magnitude for the 
hadronic amplitudes.

To close this section we emphasize that there are many interesting problems
depending on completely unknown parameters on which we have no clue, not
even a rough order of magnitude. This is for example the case in generic
new physics scenarios, such as the one of the second paper of
Ref.~\cite{UTfitpapers}. There arbitrary new physics contributions to
$B\overline B$ mixing is parametrized by a quantity
$C_B\,\exp(2i\phi_B)$; by definition, since no assumption is made on
specific new physics models, the analysis does not give any insight to
the ``natural'' scale of the parameter $C_B$. Thus one again may expect the
Bayesian treatment to suffer from instabilities when moving from, e.g, the 
$(C_B,\phi_B)$ to the $(x_B=c_B\cos\phi_B,y_B=c_B\sin\phi_B)$
parameterization 
(such instabilities are expected to increase with the number of new physics 
parameters). In other words the Bayesian approach as recommended by 
Ref.~\cite{nimportenawak} cannot be applied (although in practice it is) to 
the full generality of problems at hand.
\begin{acknowledgments}
Centre de Physique Th\'eorique is UMR 6207 du CNRS associ\'ee aux
Universit\'es d'Aix-Marseille I et
II et Universit\'e du Sud Toulon-Var; laboratoire affili\'e \`a
la FRUMAM-FR2291. 
This work was supported in part by the EU Contract No.
MRTN-CT-2006-035482, ``FLAVIAnet''.
\end{acknowledgments}

\end{document}